\documentclass[journal]{IEEEtran}
\usepackage{amsfonts}
\usepackage{graphicx}
\usepackage{amsfonts,color}
\usepackage[cmex10]{amsmath}
\usepackage{amssymb}
\usepackage{epsfig}
\usepackage[noadjust]{cite}
\usepackage{latexsym}
\usepackage{epstopdf}
\usepackage{amsmath, amsthm, amssymb}
\usepackage{verbatim}
\usepackage{subcaption}
\usepackage{amsmath}
\usepackage[utf8]{inputenc}
\usepackage{algorithmic}
\usepackage[algo2e]{algorithm2e}
\DeclareMathOperator{\Tr}{Tr}
\usepackage{algorithm}
\usepackage{enumitem}

\usepackage{soul}

\newtheorem{theorem}{Theorem}
\newtheorem{proposition}{Proposition}
\newtheorem{remark}{Remark}
\IEEEoverridecommandlockouts
\definecolor{orange}{RGB}{255,107,0}

\begin{document}
		\title{Reliable Detection of Unknown Cell-Edge Users Via Canonical Correlation Analysis}
		\author{Mohamed Salah Ibrahim,~\IEEEmembership{Student~Member,~IEEE,} and~Nicholas~D.~Sidiropoulos,~\IEEEmembership{Fellow,~IEEE}
			\thanks{Mohamed~Salah~Ibrahim,~and~Nicholas~D.~Sidiropoulos are with the Department of Electrical and Computer Engineering, University of Virginia, Charlottesville, VA, 22904 USA (e-mail (mi6cw,nikos@virginia.edu))}}		
\maketitle
\begin{abstract}
Providing reliable service to users close to the edge between cells remains a challenge in cellular systems, even as 5G deployment is around the corner. These users are subject to significant signal attenuation, which also degrades their uplink channel estimates. Even joint detection using base station (BS) cooperation often fails to reliably detect such users, due to near-far power imbalance, and channel estimation errors. Is it possible to bypass the channel estimation stage and design a detector that can reliably detect cell-edge user signals under significant near-far imbalance? This paper shows, perhaps surprisingly, that the answer is affirmative -- albeit not via traditional multiuser detection. Exploiting that cell-edge user signals are weak but {\em common} to different base stations, while cell-center users are unique to their serving BS, this paper establishes an elegant connection between cell-edge user detection and canonical correlation analysis (CCA) of the associated space-time baseband-equivalent matrices. It proves that CCA identifies the common subspace of these matrices, even under significant intra- and inter-cell interference. The resulting mixture of cell-edge user signals can subsequently be unraveled using a well-known algebraic signal processing technique. Interestingly, the proposed approach does not even require that the signals from the different base stations are synchronized -- the right synchronization can be automatically determined as well. Experimental results demonstrate that the proposed approach achieves order of magnitude BER improvements compared to `oracle' multiuser detection that assumes perfect knowledge of the cell-center user channels.     
\end{abstract}
\IEEEpeerreviewmaketitle
\section{introduction}
\IEEEPARstart{A}{t} the dawn of 5G, providing reliable high-speed service to users on the edge between cells remains a challenge that has persisted through several generations of cellular wireless systems. In 4G and legacy systems, the problem is usually tackled using aggressive power control, multiuser detection, and dynamic base station (BS) assignment / handoff~\cite{kiani2007maximizing,tolli2008cooperative}. Multiuser detection (MUD) is computationally complex (optimal MUD is NP-hard) \cite{verdu1998multiuser,proakis2001digital}, requires accurate channel estimates for all users, and while it can tolerate power imbalance, practically tractable multiuser detection does not work well in near-far scenarios, especially when the channels for the far users are not accurately known. The so-called sphere decoder (SD -- a branch-and-bound type implementation of the maximum likelihood detector) features significantly lower complexity than naive implementations at moderately high signal to noise ratios (SNRs), albeit worst-case and average complexities remain exponential  \cite{vikalo2003expected,jalden2005complexity}. Semidefinite relaxation (SDR) is a polynomial-time alternative to SD, in the low to moderate SNR regime where it yields better error rates and lower complexity than SD~\cite{tan2001application,abdi2002semidefinite}. The complexity of SDR remains high for practical implementation~\cite{5447068}. 

Minimum mean square error (MMSE)~\cite{madhow1994mmse}, and the zero-forcing (ZF -- also known as the decorrelating) detector are low-complexity linear detectors, whose performance remains far from optimal in general. ZF and MMSE detectors can be further improved  by successively canceling the strong user signals once they are decoded -- a technique referred to as successive interference cancellation (SIC), decision feedback (DF) \cite{moshavi1996multi,duel1993decorrelating}, or `turbo' (iterative) interference cancellation \cite{alexander1999iterative}. 

Although all of the aforementioned detectors have been proven successful in many applications, their detection performance is contingent on the availability of accurate channel estimates. In wireless cellular systems, accurate channel estimates may be acquired for cell-center (strong) users, however, cell-edge (weak) user signals are received at low SNR due to the inverse power law relationship between received signal power and distance. This and the intra- and inter-cell interference (particularly prominent for the cell-edge users) together induce high uncertainty in the cell-edge user channel estimates, degrading their detection performance and even leading to connection drops \cite{boudreau2009interference,himayat2010interference}. Furthermore, the frequent hand-offs of such users further complicate their situation~\cite{agarwal2014qos}. While power control~\cite{castellanos2008performance} and scheduling algorithms~\cite{chayon2017enhanced,xu2013joint} serve as two possible candidates that can considerably enhance cell-edge user detection performance, this comes at the expense of significantly reducing the rates of cell-center users. These are the ones with the best channels, so throttling their rate has a serious impact on the overall sum rate of the system.  

This begs the question whether it is possible to reliably detect cell-edge user signals without knowing their channels or sacrificing cell-center user rates? 

This paper shows that with a suitable base station ‘interferometry’ strategy inspired from machine learning, together with a well-known algebraic signal processing tool, the cell-edge user signals can be reliably decoded under mild conditions, even at low SNR and when buried under heavy intra-cell and inter-cell interference. Exploiting the fact that cell-edge user signals are {\em weak but common} to both base stations, while users close to a base station are unique to that base station, reliable detection is enabled by Canonical Correlation Analysis (CCA)~\cite{hotelling1936relations,hardoon2004canonical} – a machine learning technique that reliably estimates a common subspace using eigendecomposition, even in the presence of strong interference. 

Our approach is very different from multi-user detection using base station cooperation~\cite{khattak2008base}, as it capitalizes on CCA. CCA has been employed in several signal processing, communications, and machine learning  applications, including array processing~\cite{ge2009does}, multiple-input multiple-output (MIMO) equalization~\cite{dogandzic2002finite,via2005adaptive}, direction-of-arrival (DoA) estimation~\cite{wu1994music}, radar anti-jamming~\cite{bai2005radar} and blind source separation~\cite{li2009joint,borga2001canonical,liu2007analysis,bertrand2015distributed}, and multi-view learning~\cite{arora2014multi}, to name a few applications; but not anywhere close to our present context. Scalable algorithms for generalized (multi-view) CCA were recently developed by the authors' group~\cite{FuGCCA,FuMAXVAR,KanSUMCOR}, also incorporating various constraints.
 
\subsection{Contributions}   
This paper proposes a two-stage learning based approach that leverages base station cooperation to reliably detect cell-edge user signals without knowing their channels. The idea relies on connecting canonical correlation analysis with cell-edge user detection. In the first stage, CCA is invoked to find the common subspace of two space-time matrices, containing the baseband-equivalent signals received at two base stations. A basis for this common subspace is a mixture of the cell-edge user signals. In the second stage, this mixture is unraveled in an unsupervised fashion, using a classical algebraic technique from array signal processing, namely (R)ACMA \cite{van1997analytical}. (R)ACMA exploits constant modulus structure in the transmitted cell-edge signals, owing to digital binary/M-ary phase shift keying (BPSK or MPSK) modulation, to recover the individual cell-edge signals.  Judicious experiments demonstrate that the proposed approach works remarkably well without any power control under realistic levels of intra-cell and inter-cell interference (following the urban macro scenario from the 3GPP $38.901$ standard), delivering order of magnitude error rate  improvements compared to `oracle' multiuser detection that assumes perfect knowledge of the cell-center user channels.  Furthermore, the proposed approach does not even require that the signals from the different base stations are synchronized -- the right synchronization can be automatically determined as well. 

Beyond these compelling contributions to the particular application in cellular communications considered herein, this paper makes two notable theoretical contributions of broader interest.  First, it proves that CCA identifies the common subspace between two matrices, under a rather general (and purely deterministic) linear generative model. Second, it includes a performance analysis which shows that CCA works even in the non-ideal case where there is background noise and `leakage' of the individual components to the other matrix view --  e.g., the case where there is thermal noise and realistic adjacent-cell interference from non-cell-edge users that cannot be neglected, in the context of our application herein.

The overall complexity of the proposed method depends on the cost incurred in solving CCA and RACMA. Fortunately, both admit relatively simple algebraic solution via eigenvalue decomposition~\cite{hardoon2004canonical,van1997analytical}. This renders the overall approach computationally efficient even when the base station is equipped with a large number of antennas and is serving a large number of users.

A preliminary version of part of this work was presented at IEEE SPAWC 2019~\cite{salah2019}. This journal version includes performance analysis that  was missing from \cite{salah2019}, a new section showing how the common cell-edge signals can be used to synchronize the signals from the two base stations even if they were asynchronously acquired (thus alleviating the need for symbol-level synchronization), and a new comprehensive suite of experiments to demonstrate the superior performance of the proposed method in more practical scenarios. 

\subsection{Outline of the Paper} 	
The rest of this paper is organized as follows. After a succinct introduction to CCA in Section \ref{RCCA}, Section \ref{Model} describes the system model and gives a brief review on cell-edge user detection. The proposed detector is presented in Section \ref{PD}, while Section \ref{SYNC} explains how our detector can be used to resolve symbol synchronization between the two base stations. Simulation results are provided in Section \ref{Simu}, and conclusions are drawn in Section \ref{Conc}. Long proofs and derivations are relegated to the Appendix. 





%
\subsection{Notation}
In this work, we use upper and lower case bold letters to denote matrices and column vectors, respectively. For any general matrix $ {\bf G} $, we use ${\bf G}^T$,  ${\bf G}^H$, ${\bf G}^{-1}$, ${\bf G}^{\dagger}$ and $ \Tr({\bf G}) $ to denote the transpose, the conjugate-transpose, the inverse, the pseudo-inverse, and the trace of $ {\bf G} $, respectively. Scalars are represented in the normal face, while calligraphic font is reserved for sets. $ \lVert.\rVert_2 $ and $  \lVert.\rVert_F  $ denote the $ \ell_2 $-norm and the Frobenius norm, respectively. Finally, $ {\bf I}_N $ and $ {\bf 0}_{N \times M} $ denote the $ N \times N $ identity matrix and the $ N \times M $ zero matrix, respectively.

\section{Canonical Correlation Analysis}\label{RCCA}
Consider $ T $ samples of the pair $ ({\bf y}_1,{\bf y}_2) $, where $ {\bf y}_1 \in \mathbb{R}^{M_1} $ and $ {\bf y}_2 \in \mathbb{R}^{M_2} $ are two ``views'' of the same entity. For example, ${\bf y}_1$ could contain a set of economic indicators, while ${\bf y}_2$ could contain crime, corruption, or social welfare data corresponding to the same country or municipality, and we have data for $T$ countries or municipalities. Or, ${\bf y}_1$ could be the electroencephalogram (EEG) of a person and ${\bf y}_2$ could be the voxels of a functional magnetic resonance (fMRI) scan; or ${\bf y}_1$ could be a person's consumer record, while ${\bf y}_2$ could reflect his/her social network connections, and we have data for $T$ people. We are interested in discovering what is common between these two views of the same set of entities. Is there a particular `latent' factor that affects both the economy and crime, for example? Towards this end, we would like to derive `meta-variables', one from each view, which are strongly correlated with each other. How can we do this?

Let  $ {\bf y}_1[t] $ and $ {\bf y}_2[t] $ denote the $ t $-th observation of $ {\bf y}_1 $ and $ {\bf y}_2 $, respectively, corresponding to the $t$-th entity, for $ t \in \{1,\cdots,T\} $. Assume that both $ {\bf y}_1 $ and $ {\bf y}_2 $ are zero-mean, otherwise the sample mean can be subtracted as a pre-processing step. In its simplest form, CCA seeks to find a pair of linear combinations of the variables in the two respective views which are highly correlated to each other -- ideally, perfectly correlated. Mathematically, CCA seeks to weight vectors $ {\bf q}_1 \in \mathbb{R}^{M_1} $ and $ {\bf q}_2 \in \mathbb{R}^{M_2} $ such that the correlation coefficient between $ {\bf Y}_1^T{\bf q}_1  $ and $ {\bf Y}_2^T{\bf q}_2  $ is maximized, where $ {\bf Y}_\ell := [{\bf y}_\ell[1],\cdots,{\bf y}_\ell[T]]  \in \mathbb{R}^{M_\ell \times T}$ and $ \ell \in \{1,2\} $. In an optimization framework, this can be expressed as
\begin{equation}\label{R1}
	\underset{{\bf q}_1,{\bf q}_2}{\max}~\frac{{\bf q}_1^T{\bf Y}_1{\bf Y}_2^T{\bf q}_2}{\sqrt{{\bf q}_1^T{\bf Y}_1{\bf Y}_1^T{\bf q}_1}\sqrt{{\bf q}_2^T{\bf Y}_2{\bf Y}_2^T{\bf q}_2}}
\end{equation}
Let $ {\bf R}_{{\bf y}_\ell{\bf y}_\ell} := \frac{1}{T}{\bf Y}_\ell{\bf Y}^T_\ell $ and  $ {\bf R}_{{\bf y}_1{\bf y}_2} := \frac{1}{T}{\bf Y}_1{\bf Y}^T_2 $ denote the sample auto-correlation matrix of $ {\bf y}_\ell $ and the sample cross-correlation matrix of $ {\bf y}_1 $ and $ {\bf y}_2 $, respectively. Then,~\eqref{R1} can be equivalently written as   
\begin{subequations}\label{R2}
	\begin{align}
	&\underset{{\bf q}_1,{\bf q}_2}{\max}~{\bf q}_1^T  {\bf R}_{{\bf y}_1{\bf y}_2}{\bf q}_2 \\
	& \text{s.t.} \quad 	{\bf q}_\ell^T {\bf R}_{{\bf y}_\ell{\bf y}_\ell}{\bf q}_\ell = 1,~\ell = 1,2 \label{R2b}
	\end{align} 
\end{subequations}
where the constraints in~\eqref{R2b} arise from the fact that the objective of~\eqref{R1} is not affected by re-scaling $ {\bf q}_1  $ and$ / $or $ {\bf q}_2 $. Using the Lagrange duality theorem, a solution of~\eqref{R2} can be provided in closed-form. The Lagrangian of~\eqref{R2} is
\begin{equation}\label{R3}
 \mathcal{L}({\bf q}_1,{\bf q}_2,\lambda_1,\lambda_2) = {\bf q}_1^T  {\bf R}_{{\bf y}_1{\bf y}_2}{\bf q}_2 - \sum_{\ell = 1}^{2}\frac{\lambda_\ell}{2}({\bf q}_\ell^T {\bf R}_{{\bf y}_\ell{\bf y}_\ell}{\bf q}_\ell - 1)
\end{equation}
By taking the derivatives with respect to $ {\bf q}_1 $ and $ {\bf q}_2 $, we obtain
\begin{equation}\label{R4}
 \frac{\partial \mathcal{L}}{\partial {\bf q}_1} =  {\bf R}_{{\bf y}_1{\bf y}_2}{\bf q}_2 - \lambda_1{\bf R}_{{\bf y}_1{\bf y}_1}{\bf q}_1 = 0
\end{equation}
\begin{equation}\label{R5}
\frac{\partial \mathcal{L}}{\partial {\bf q}_2} =  {\bf R}_{{\bf y}_2{\bf y}_1}{\bf q}_1 - \lambda_2{\bf R}_{{\bf y}_2{\bf y}_2}{\bf q}_2 = 0
\end{equation}
By left multiplying~\eqref{R4} and~\eqref{R5} with $ {\bf q}_1^T $ and $ {\bf q}_2^T $, respectively, we have
\begin{equation}\label{R6}
    {\bf q}_1^T{\bf R}_{{\bf y}_1{\bf y}_2}{\bf q}_2 = \lambda_1{\bf q}_1^T{\bf R}_{{\bf y}_1{\bf y}_1}{\bf q}_1 
\end{equation}
\begin{equation}\label{R7}
  {\bf q}_2^T{\bf R}_{{\bf y}_2{\bf y}_1}{\bf q}_1 = \lambda_2{\bf q}_2^T{\bf R}_{{\bf y}_2{\bf y}_2}{\bf q}_2
\end{equation}
which together with the constraints in~\eqref{R2b} imply that $ \lambda_1 = \lambda_2 = \lambda $. By assuming that the matrix ${\bf R}_{{\bf y}_2{\bf y}_2} $ is invertible, the optimal solution, $ {\bf q}_2^{*} $, of~\eqref{R5} is given by
\begin{equation}\label{R8}
          {\bf q}_2^{*} = \frac{1}{\lambda}{\bf R}_{{\bf y}_2{\bf y}_2}^{-1}{\bf R}_{{\bf y}_2{\bf y}_1}{\bf q}_1^*
\end{equation}
Then by substituting in~\eqref{R4},  the optimal solution, $ {\bf q}_1^{*} $, can be obtained by solving the following generalized eigenvalue problem
\begin{equation}\label{R9}
{\bf R}_{{\bf y}_1{\bf y}_2}{\bf R}_{{\bf y}_2{\bf y}_2}^{-1}{\bf R}_{{\bf y}_2{\bf y}_1}{\bf q}_1 = \lambda^2{\bf R}_{{\bf y}_1{\bf y}_1}{\bf q}_1
\end{equation}
It can be easily seen from~\eqref{R4} that the maximum eigenvalue $ \lambda^{*} $ of~\eqref{R9} is nothing but the square of the correlation coefficient, $ \rho_1 $, associated with the optimal canonical pair $ ({\bf q}_1^*,{\bf q}_2^*) $.

Considering the generalization to $ N \leq \min(M_1,M_2) $ canonical pairs, $ \{({\bf q}_1[n],{\bf q}_2[n])\}_{n = 1}^{N} $. After identifying $ {\bf q}^*_1[1] = {\bf q}_1^* $  and $ {\bf q}^*_2[1] = {\bf q}_2^* $, we can iteratively solve the following problem
\begin{subequations}\label{R10}
	\begin{align}
     & 	\underset{{\bf q}_1[n],{\bf q}_2[n]}{\max}  &&  {\bf q}_1^T[n]  {\bf R}_{{\bf y}_1{\bf y}_2}{\bf q}_2[n] \\
	& \text{s.t.}  &&	{\bf q}_\ell^T[n] {\bf R}_{{\bf y}_\ell{\bf y}_\ell}{\bf q}_\ell[n] = 1,~\ell = 1,2 \\
	&&& {\bf q}_\ell^T[n] {\bf R}_{{\bf y}_\ell{\bf y}_\ell}{\bf q}_\ell[j] = 0,~ j = 1,\cdots,n-1
	\end{align} 
\end{subequations} 
for $ n = \{2,\cdots,N \}$. Instead of solving $ N $ sub-problems of type~\eqref{R10}, we can instead solve one joint problem. Let us stack the vectors $ \{{\bf q}_\ell[n]\}_{n=1}^{N} $ in the matrix $ {\bf Q}_\ell \in \mathbb{R}^{M_\ell \times N} $, for $ \ell \in \{1,2\} $, and rewrite~\eqref{R10} in the following compact form
\begin{subequations}\label{R11}
	\begin{align}
	&\underset{{\bf Q}_1,{\bf Q}_2}{\max} \Tr({\bf Q}_1^T{\bf R}_{{\bf y}_1{\bf y}_2}{\bf Q}_2) \\
	& \text{s.t.} \quad ~	{\bf Q}_\ell^T {\bf R}_{{\bf y}_\ell{\bf y}_\ell}{\bf Q}_\ell = 1,~\ell =1,2
	\end{align} 
\end{subequations} 
which yields simultaneously multiple canonical pairs. Following the same procedures for solving~\eqref{R2}, it can be shown that the optimal solution $ {\bf Q}_1^*$ should satisfy the following generalized eigenvalue equation
\begin{equation}\label{R12}
    {\bf R}_{{\bf y}_1{\bf y}_2}{\bf R}_{{\bf y}_2{\bf y}_2}^{-1}{\bf R}_{{\bf y}_2{\bf y}_1}{\bf Q}_1 = {\bf R}_{{\bf y}_1{\bf y}_1}{\bf Q}_1\Lambda^2
\end{equation}
where $ \Lambda = \text{Diag}([\rho_1,\cdots,\rho_N]) $ with $ \rho_\ell $ be the $ \ell $-th correlation coefficient associated with the $ \ell $-th canonical pair, for $ \ell = \{1,\cdots,N\} $. Note that the optimal solution $ {\bf Q}_2^*$ can be directly obtained from~\eqref{R8} after solving~\eqref{R12}. 

The two-view CCA problem in~\eqref{R11} can be equivalently formulated as a distance minimization between the low dimensional representations $ {\bf Y}_1^T{\bf Q}_1 $ and $ {\bf Y}_2^T{\bf Q}_2 $~\cite{carroll1968generalization,hardoon2004canonical}, where the distance is measured by the Frobenius norm, i.e.,
\begin{subequations}\label{R13}
	\begin{align}
	&\underset{{\bf Q}_1,{\bf Q}_2}{\min}~ \| {\bf Y}_1^T{\bf Q}_1 - {\bf Y}_2^T{\bf Q}_2 \|^2_F\\
	& \text{s.t.} \quad~ {\bf Q}_\ell^T{\bf Y}_\ell{\bf Y}_\ell^T{\bf Q}_\ell = \textbf{\textit{I}},~~ \ell = 1,2 
	\end{align}
\end{subequations}
Note that by expanding the objective in~\eqref{R13}, the equivalence between~\eqref{R12} and~\eqref{R13} can be readily verified. In what follows, we will see how the CCA approach can be utilized to handle the problem of cell-edge user detection in a multi-cell multi-user system.
\section{Problem Statement}\label{Model}
\subsection{System Model}
\begin{figure}[!t]
	\centering
	\includegraphics[width=3in]{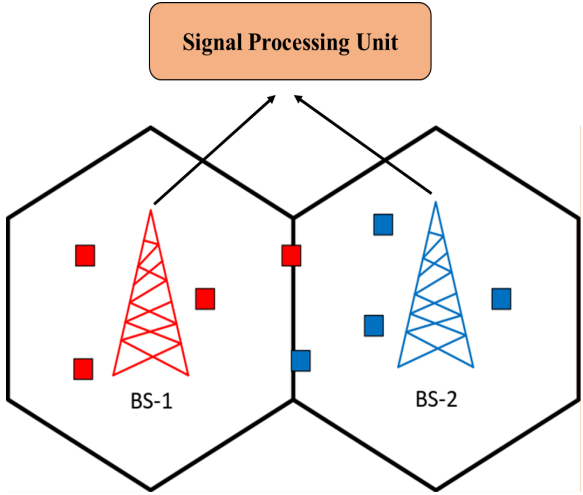}
	\caption{System Model}
	\label{SM1}
\end{figure}

Consider a multi-cell multi-user MIMO system comprising two hexagonal cells with a single base station (BS) located at the center of each cell, as shown in~Figure~\ref{SM1}. The $ \ell $-th BS is equipped with $ M_\ell $ antennas and serves $ K_\ell $ single-antenna users, for $ \ell \in \{1,2\} $. Let $ K_e = K_{e_1}+K_{e_2}$ denote the total number of cell-edge users located around the common edge between the two cells, where $ K_{e_\ell} < K_\ell$ represents the number of cell-edge users served by the $ \ell $-th BS. Let  $  {\bf h}_{\ell kj}  $  model path-loss and small scale fading between the $ k $-th user in the $ j $-th cell and the $ \ell $-th BS, given by
\begin{equation}\label{PS1}
{\bf h}_{\ell kj} = \sqrt{\alpha_{\ell kj}}{\bf g}_{\ell kj}
\end{equation}
where $ {\bf g}_{\ell kj} \in \mathbb{C}^{M_\ell \times 1}$ represents the small scale fading between user $ k $ in cell $ j $ and BS $ \ell $, while $ \alpha_{\ell kj} \in \mathbb{R}$ models the large scale fading that accounts for the path loss between BS $ \ell $ and user $ k $ in cell $ j $. Throughout this work, it is assumed that the uplink channel vectors $  {\bf h}_{\ell kj}  $ for the cell-edge users are not \textit{a priori} known at any BS.

\subsection{Uplink transmission}
Consider uplink transmission from the users to the BSs where each user aims at transmitting its data to its serving BS. We assume that all users access the same channel without any (sub-)channel allocation or coordination mechanism, thereby creating intra- and inter-cell interference. Define $ {\bf s}_{kj} \in \mathbb{R}^{T \times 1}$  to be the vector containing symbols transmitted by the $ k $-th user in cell $ j $, where each entry of $ {\bf s}_{kj}  $  belongs to the finite alphabet $ \Omega = \{\pm 1\}$ (our approach works for general PSK and other alphabets, with some variations in the second stage). The received signal, $ {\bf Y}_{\ell} \in \mathbb{C}^{M_\ell \times T} $, at the $ \ell $-th BS can be expressed as
\begin{equation}\label{UT1}
{\bf Y}_\ell = \sum_{j = 1}^{2}\sum_{k = 1}^{K_j} \sqrt{\beta}_{kj}{\bf h}_{\ell kj}{\bf s}_{kj}^T + {\bf W}_{\ell}
\end{equation}
where $ {\bf h}_{\ell kj} \in \mathbb{C}^{M_\ell \times 1 }$ is the uplink channel response vector defined in~\eqref{PS1}, $ {\mathbf W}_\ell \in \mathbb{C}^{M_\ell \times T}$ contains independent identically distributed (i.i.d.) complex Gaussian entries of zero mean and variance $ \sigma^2 $, and $ \beta_{kj} $ represents the transmit power of the $ k$-th user in the $ j $-th cell. 

Throughout this paper, we assume that each BS forwards its received signal to a central signal processing unit (CSPU). Although BSs cooperation has been considered before for the sake of mitigating inter-cell interference~\cite{mayer2006turbo}, cooperation here is assumed for a very different purpose. That is, we leverage the joint processing of the BSs signals at the CSPU to provide reliable detection of cell-edge user signals at low SNR, without knowledge of their channels. Furthermore, in contrast to prior cooperation strategies that assume perfect synchronization of the received signals from different BSs \cite{khattak2008base,hoymann2009distributed}, this work deals with BS asynchrony as well, rendering the approach more practical. Specifically, it will be shown in Section~\ref{SYNC} how the proposed method can detect the cell-edge user signals even if there exists a time delay between $ {\bf Y}_1 $ and $ {\bf Y}_2 $.
\subsection{Cell-Edge User Detection}
Let us denote the cell-edge user signals received at BS $\ell$ by $ {\bf S}_c \in \mathbb{R}^{T \times K_e}  $ (where the subscript {\it c} stands for {\it common}), and those of the cell-center signals received at BS $\ell$ as $ {\bf S}_{p_\ell} \in \mathbb{R}^{T \times (K_\ell-K_{e_\ell})}  $ (where the subscript {\it p} stands for {\it private}).  Furthermore, let $ \widetilde{{\bf W}}_\ell  $ represent the noise at the $ \ell $-th BS plus the inter-cell interference caused by the cell-center users in cell $ j $, where $ j \neq \ell $. Therefore,~\eqref{PS1} can be expressed as follows
\begin{equation}\label{CE}
{\bf Y}_\ell = {\bf H}_{\ell p_\ell}{\bf S}_{p_{\ell}}^T + {\bf H}_{\ell c}{\bf S}^T_c + \widetilde{\bf W}_\ell   
\end{equation}
where the matrices $ {\bf H}_{\ell c} \in \mathbb{C}^{M_\ell \times K_e }$ and $ {\bf H}_{\ell p_\ell} \in \mathbb{C}^{M_\ell \times (K_\ell-K_{e_\ell}) }$ hold on their columns all the channel vectors from  cell-edge users to the $ \ell $-th BS, and the channel vectors from cell-center users to their serving BS, respectively. Moreover, absorb the transmitted signal power, $ \beta_{kj} $, of the $ k $-th user in the $ j $-th cell in its respective channel vectors, $ \forall~ k,j $.  

In general, to guarantee reliable detection performance for each cell-edge user, its serving BS requires accurate knowledge about its channel state information (CSI)~\cite{bjornson2016massive,li2015multi,ngo2012performance}. However, due to the fact that cell-edge user signals are often received intermittently at very low signal to interference plus noise ratio (SINR) and SNR, their channel estimates are inaccurate \cite{boudreau2009interference,himayat2010interference}.

One possible approach to detect cell-edge user signals is to apply zero-forcing successive interference cancellation (ZF-SIC)~\cite{moshavi1996multi}, which is based on successively removing the cell-center (strong) user signals once they are detected using ZF. Applying SIC after ZF improves the detection performance of the cell edge user signals as it (ideally) cancels the strong interference that stems from the transmissions of cell-center users, i.e., intra-cell interference. However, cell-center user detection is imperfect, which can lead to error propagation, and in-cell SIC does not address the inter-cell interference, which is particularly prominent for the cell-edge users.  In the absence of power control~\cite{castellanos2008performance} and/or scheduling~\cite{chayon2017enhanced} , cell-edge user detection performance is severely affected by the intra-cell interference from cell-center users. In what follows, we present a novel \textit{blind} detector that can reliably decode cell-edge user signals at low received SNR and without knowing their channels. 


\section{Cell-Edge User Detection Via CCA} \label{PD}
In this section, it is assumed that the base stations signals are perfectly synchronized at the CSPU. We will explain how to deal with asynchrony later. The goal of the proposed detector is to decode cell-edge user signals $ {\bf S}_c $ from the received signals $ {\bf Y}_1$ and ${\bf Y}_2 $. As a pre-processing step, the signals are transformed to the real domain by forming the matrix $ \overline{\bf Y}_\ell := [{\bf Y}_\ell^{(r)};{\bf Y}_\ell^{(i)}]  \in \mathbb{R}^{2M_\ell \times T}$, where  $ {\bf Y}_\ell^{(r)} = {\rm I\!Re}\{{\bf Y}_\ell\} $ and $ {\bf Y}_\ell^{(i)} = {\rm I\!Im}\{{\bf Y}_\ell\}  $ represent the real and imaginary components of the $ \ell $-th BS signal.   
Similarly, denote by $ \overline{\bf A}_{\ell p_\ell} := [{\bf H}_{\ell p_\ell}^{(r)};{\bf H}_{\ell p_\ell}^{(i)}]  \in \mathbb{R}^{2M_\ell \times( K_\ell - K_{e_\ell})}$, $ \overline{\bf A}_{\ell c} := [{\bf H}_{\ell c}^{(r)};{\bf H}_{\ell c}^{(i)}]  \in \mathbb{R}^{2M_\ell \times K_e}$ and $ \overline{\bf W} = [\widetilde{\bf W}^{(r)}_\ell;\widetilde{\bf W}^{(i)}_\ell] \in \mathbb{R}^{2M_\ell \times T}$. Therefore, \eqref{CE} can be equivalently expressed as
\begin{equation}\label{MV2}
\overline{\bf Y}_\ell = \overline{\bf A}_{\ell p_\ell}{\bf S}_{p_{\ell}}^T + \overline{\bf A}_{\ell c}{\bf S}^T_c + \overline{\bf W}_\ell.   
\end{equation}

In what follows, the two-view CCA formulation in~\eqref{R13} is exploited to estimate the subspace containing the cell-edge user signals. For the sake of brevity, we refer to this subspace as the \textit{common subspace}. Define the two matrices $ \overline{\bf Q}_1 \in \mathbb{R}^{2M_1 \times N}$ and $ \overline{\bf Q}_2 \in \mathbb{R}^{2M_2 \times N} $,  where  the $ n $-th column of $ \overline{\bf Q}_\ell $ represents the $ n $-th canonical component of view $ \overline{\bf Y}_\ell $, for $n \in \{1,\cdots,N\}$. The number of components (pairs) extracted, ($ N $), depends on the minimum value of the correlation coefficient that needs to be considered. 

An alternative formulation of~\eqref{R13} is to search for an orthogonal representation $ {\bf G} \in \mathbb{R}^{T \times N}$ that is maximally correlated after the linear projections of $ \overline{\bf Y}_1 $ and $ \overline{\bf Y}_2 $ on $ \overline{\bf Q}_1 $ and $ \overline{\bf Q}_2 $, respectively. This can be written as
\begin{subequations}\label{MV4}
	\begin{align}
	&\underset{\overline{\bf Q}_1,\overline{\bf Q}_2,{\bf G}}{\min}~ \sum_{\ell = 1}^{2}\| \overline{\bf Y}_\ell^T\overline{\bf Q}_\ell - {\bf G}\|^2_F\\
	& \text{s.t.} \quad~~~~ {\bf G}^T{\bf G} = \textbf{\textit{I}}\label{MV4b}
	\end{align}
\end{subequations}
Problem~\eqref{MV4} is known as the maximum-variance (MAX-VAR) formulation of CCA~\cite{horst1961generalized,carroll1968generalization}, and, in the case of two views considered here, it is equivalent to~\eqref{R13} in the sense that both problems yield the same solution $ \overline{\bf Q}^{*}_\ell $. In this section, we focus on the formulation in \eqref{MV4} as it facilitates our proof. 

Assume that we are interested in the first $ K_e $ canonical components of the matrices $ \overline{\bf Q}_1 $ and $ \overline{\bf Q}_2 $, i.e., $ N = K_e $. We have the following result.  
\begin{theorem}
	In the noiseless case, if matrix $ {\bf B} := [{\bf S}_c,{\bf S}_{p_1},{\bf S}_{p_2}] \in \mathbb{R}^{T \times (K_1 + K_2)} $ is full column rank, and  $\overline{\bf A}_\ell = [\overline{\bf A}_{\ell c},\overline{\bf A}_{\ell p_\ell}] \in \mathbb{R}^{2M_\ell \times (K_e + K_\ell - K_{e_\ell})}$ is full column rank for $\ell \in \left\{1,2\right\}$, then the optimal solution $ {\bf G}^{\star} $ of problem~\eqref{MV4} is given by $ {\bf G}^{\star} = {\bf S}_c {\bf P} $, where $ {\bf P} $ is a $ K_e \times K_e $ non-singular matrix. 
\end{theorem}
\begin{remark}
	The full column rank condition on ${\bf B}$ requires $T~ \text{greater than or equal to} ~(K_1 + K_2)$, and the transmitted sequences from the different users to be linearly independent. For finite-alphabet signals, this occurs with very high probability for modest $T$, since the different user transmissions are independent. The more restrictive condition is full column rank of $\overline{\bf A}_\ell$, which relates the number of base station antennas and signals impinging on each base station. We thus need two times the number of antennas in each base station to be greater than or equal to the number of users assigned to that base station, plus any cell-edge users assigned to the other base station. Other than this dimensionality constraint though, if the channel vectors are drawn from a jointly continuous distribution, the latter condition will be satisfied with probability one. 
\end{remark}
\begin{proof}
	First, let us start with the single cell-edge user case, i.e., $ K_e = 1 $ and each of $ {\bf S}_c, {\bf G}  $ and $ {\bf Q}_\ell $ is a vector. In such setting~\eqref{MV4} reduces to the following 
	\begin{subequations}\label{MV5}
		\begin{align}
		&\underset{\overline{\bf q}_1,\overline{\bf q}_2,{\bf g}}{\min}~ \sum_{\ell = 1}^{2}\| \overline{\bf Y}_\ell^T\overline{\bf q}_\ell - {\bf g}\|^2_2\\
		& \text{s.t.} \quad \| {\bf g} \|^2_2= 1
		\end{align}
	\end{subequations}
	To solve the above problem, we need to find $ (\overline{\bf q}_1^{*},\overline{\bf q}_2^{*},{\bf g}^{*}) $ that can together attain a zero-cost. In other words, we need the following two conditions to be satisfied simultaneously
	\begin{subequations}\label{MV6}
		\begin{align}
		&   \overline{\bf Y}_1^T\overline{\bf q}_1 =  {\bf g}\label{MV6a}\\
		&   \overline{\bf Y}_2^T\overline{\bf q}_2 =  {\bf g}\label{MV6b} 
		\end{align} 
	\end{subequations}
	Without loss of generality, we can let $ \overline{\bf q}_\ell = \overline{\bf A}_\ell(\overline{\bf A}_\ell^T\overline{\bf A}_\ell)^{-1}{\bf u}_\ell$, where ${\bf u}_\ell $ is any vector in  $ \mathbb{R}^{K_e + K_\ell - K_{e_\ell}} $. The reason is that we can always decompose $ \overline{\bf q}_\ell$ into a component in the subspace spanned by $\overline{\bf A}_\ell$ and one orthogonal to it. The latter is annihilated anyway after multiplication with $\overline{\bf A}^T_\ell$.  Substituting in~\eqref{MV6a} and~\eqref{MV6b} and taking their difference, we obtain  
	\begin{equation}\label{MV7}
	{\bf B}{\bf u} = {\bf 0},
	\end{equation}
	where $ {\bf B} = [{\bf s}_c,{\bf S}_{p_1},{\bf S}_{p_2}] \in \mathbb{R}^{T \times (K_1+K_2)} $ and $ {\bf u}=[{\bf u}_1(1) - {\bf u}_2(1),{\bf u}_{1}(2:\text{end}),-{\bf u}_{2}(2:\text{end})]^T \in \mathbb{R}^{(K_1+K_2)} $, where ${\bf u}_{1}(2:\text{end})$ is the vector containing all except the first element of ${\bf u}$. It can be easily seen that if ${\bf B} $ is full column rank, then $ {\bf u} = {\bf 0}_{(K_1 + K_2) \times 1} $ is the only possible  solution of~\eqref{MV7}. This means that $ {\bf u}_1 = {\bf u}_2 = c{\bf e}_1$, where $ c $ is any constant and $ {\bf e}_1 $ is the first column of the identity matrix. Consequently, from~\eqref{MV6}, $ {\bf g}^{\star} =  \alpha{\bf s}_c/\|{\bf s}_c\|_2 $, with $ \alpha = \pm 1$, will be the only possible solution for problem~\eqref{MV5}.
	
	The generalization to $ K_e > 1$ now follows naturally. Letting $ \overline{\bf Q}_\ell  =  \overline{\bf A}_\ell(\overline{\bf A}_\ell^T\overline{\bf A}_\ell)^{-1} {\bf U}_\ell$, and defining 
	\[
	{\bf U} := \left[
	\begin{array}{ll}
	& {\bf U}_1(1:K_e,:) - {\bf U}_2(1:K_e,:)\\
	& {\bf U}_{1}(K_e+1:\text{end},:)\\
	- & {\bf U}_{2}(K_e+1:\text{end},:)
	\end{array}
	\right] \in \mathbb{R}^{(K_1+K_2) \times K_e},
	\]
	where ${\bf U}_1(1:K_e,:)$ means rows $1$ to $K_e$ and all columns of ${\bf U}_1$, we obtain
	\begin{equation}\label{MVN}
	{\bf B}{\bf U} = {\bf 0},
	\end{equation}
	and when ${\bf B}$ is full column rank the solution is unique: ${\bf U} = {\bf 0}$,  and therefore ${\bf U}_1(1:K_e,:)={\bf U}_2(1:K_e,:)=:{\bf P}$, ${\bf U}_{1}(K_e+1:\text{end},:)={\bf 0}$, ${\bf U}_{2}(K_e+1:\text{end},:)={\bf 0}$, and therefore  $ {\bf G}^{\star} = {\bf S}_c{\bf P}  $, where $ {\bf P} $ is $ K_e \times K_e $ non-singular such that the orthonormality constraint~\eqref{MV4b} is satisfied. Note that if the signals themselves are (approximately) orthogonal, then ${\bf P}$ will be orthogonal as well, which helps with the next (RACMA) stage.   
\end{proof}

The next step is to extract the cell-edge user sequences $ {\bf S}_c $ from $ {\bf G}^{\star} =  {\bf S}_c{\bf P}$.  
This problem can be viewed as a bi-linear factorization of the matrix  $ {\bf G}^{\star} $ to its factors $ {\bf P} $ and $ {\bf S}_c $ under the constraint that the entries of $ {\bf S}_c $ belong to the finite alphabet $ \Omega = {\pm 1}  $. This can be mathematically posed as an optimization problem as follows
\begin{subequations}\label{CSI2}
	\begin{align}
	&\underset{\overline{\bf S}_c,\overline{\bf P}}{\min}~ \| {\bf G}^{\star} - \overline{{\bf S}}_c\overline{\bf P} \|_F^2 \\
	&\text{s.t.} \quad \overline{{\bf S}}_c(i,j) \in \Omega
	\end{align}
\end{subequations}
In \cite{van1997analytical}, van der Veen proposed an algebraic algorithm called Real Analytical Constant Modulus Algorithm (RACMA) for this problem. RACMA does not claim to optimally solve \eqref{CSI2}, which is NP-hard even if  $ {\bf P} $ is known. Instead, RACMA assumes that noise is small, and reduces \eqref{CSI2} to a generalized eigenvalue problem. The solution is subject to sign and user permutation ambiguity. This means that the original $ {\bf S}_c $ can be identified up to permutations and column-wise (user) scaling by $ \pm 1 $. From the practical point of view, each user has its unique identification sequence, so once the users signals are received correctly each BS can identify each user signal (and sign) via correlation with the identification sequence.  

The following Algorithm describes the two-step procedure for cell-edge users detection via CCA followed by RACMA.
\begin{algorithm} 
	\textbf{Input}: $ \overline{\bf Y}_1,  \overline{\bf Y}_2 $ 
	\begin{enumerate}
		\item Solve problem~\eqref{R13} for $ \overline{\bf Q }_\ell$ as explained in Section~\ref{RCCA}
		\item Compute $ {\bf G}_\ell = \overline{\bf Y}_\ell^T\overline{\bf Q}_\ell  \in \mathbb{R}^{T \times K_e}$ 
		\item Construct $ {\bf G} = [{\bf G}_1;{\bf G}_2] \in \mathbb{R}^{2T \times K_e} $ and pass it to RACMA
		\item Compute the BER of cell-edge users by comparing the output of RACMA with the original cell-edge user transmitted sequences
	\end{enumerate}
	\caption{CCA for Cell-Edge User Detection}
	\label{A1}
\end{algorithm}

Notice that the second step in Algorithm 1 stems out from the fact that the zero-cost solution of problem~\eqref{MV4} is not guaranteed in the noisy case, and therefore, $ \overline{\bf Y}^T_1\overline{\bf Q}_1 $ is not equal to  $ \overline{\bf Y}^T_2\overline{\bf Q}_2 $ in general. Then, it turns out that feeding RACMA with both $ \overline{\bf Y}^T_1\overline{\bf Q}_1 $ and  $ \overline{\bf Y}^T_2\overline{\bf Q}_2 $ simultaneously results in much better BER as we will see in Section~\ref{Simu}.

The overall complexity of the proposed method comes from solving problems~\eqref{R13} and~\eqref{CSI2}. Fortunately, similar to ~\eqref{R13}, \eqref{CSI2} also admits simple algebraic solution via eigenvalue decomposition~\cite{van1997analytical}. This means that our end-to-end method requires solving two eigenvalue problems, i.e., the overall complexity is of $ O(M^3) $, with $ M = \max\{M_1,M_2\} $.   

It is important to emphasize that, in the noisy case and under inter-cell interference (i.e., users close to base station B can be overheard at base station A), it turns out that our method can still identify the common subspace, even at low SNR values. 
In order to show this, we follow a very different path from that described in the proof of Theorem 1. In particular, we have the following result.
\begin{proposition}
In the noisy and inter-cell interference case, if  $ \frac{1}{T}{\bf B}^T{\bf B} \approx {\bf I}  $ and  $ {\bf H}_\ell^H{\bf H}_\ell \approx {\bf I}  $, where $ {\bf H}_1 := [{\bf H}_{1 c},{\bf H}_{1 p_1},{\bf H}_{1  p_2}] \in \mathbb{C}^{M_1 \times (K_1 + K_2)}$ and $ {\bf H}_2 := [{\bf H}_{2 c},{\bf H}_{2 p_1},{\bf H}_{2 p_2}] \in \mathbb{C}^{M_2 \times (K_1 + K_2)}$, then under certain conditions on the relative SNRs of cell-center and cell-edge users (see the Appendix), the optimal solution $ {\bf Q}_\ell^{\star} $ of problem~\eqref{R13} is given by $ {\bf Q}_\ell^{\star} = {\bf H}_\ell{\bf V}{\bf M}_\ell $, where $ {\bf V} $ contains the first $ K_e $ columns of an $ (K_1 +K_2) \times (K_1 + K_2)$ identity matrix, and ${\bf M}_\ell $ is a $ K_e \times  K_e  $ non-singular matrix. 
\end{proposition}
\begin{remark}
The approximate semi-$ \perp $ constraint on the matrix $ {\bf B} $ posits that the transmitted sequences of different users are approximately orthogonal. For binary signals, this occurs with high probability for large enough $ T $, since the user transmissions are independent.  On the other hand, the approximate orthonormality constraint on $ {\bf H}_\ell $ requires the number of base station antennas to be greater than the total number of users assigned to both base stations and that the entries of $ {\bf H}_\ell $ to be drawn from a zero-mean complex Gaussian distribution with variance $ 1/M_\ell $. The dimensionality constraint on the number of antennas is supported by massive MIMO technology that aims to equip base stations with hundreds of antennas.
\end{remark}
\begin{proof}
	The proof is relegated to Appendix A.
\end{proof} 

\section{Synchronization}\label{SYNC}
In Section~\ref{PD}, we proposed a learning-based approach that can identify cell-edge user signals. However, this was under the assumption that the received signals from both BSs are perfectly synchronized at the CSPU. One natural question that can be posed is what if there exists a time delay $ \tau_d $ between $ \overline{\bf Y}_1 $ and $ \overline{\bf Y}_2 $ at the CSPU. It turns out that our proposed method not only can recover cell-edge user signals in the synchronized case, it can even detect the time difference, $ \tau_d $, between the two signals, re-synchronize the signals and then decode
 them as explained in Section~\ref{PD}.

Assume that the CSPU has received two long sequences $ \widetilde{\bf Y}_1  \in \mathbb{R}^{2M_1 \times \widetilde{T}}$ and $ \widetilde{\bf Y}_2 \in \mathbb{R}^{2M_2 \times \widetilde{T}}$, where $ \widetilde{T} > T $, and that the sequence length $ T $ is known or has been estimated~\cite{wax1985detection} at the CSPU. The goal is to find the correct delay between the signals $ \widetilde{\bf Y}_1 $ and $ \widetilde{\bf Y}_2 $ so that we can extract the desired signal $ \overline{\bf Y}_\ell $ from $ \widetilde{\bf Y}_\ell $, and then apply Algorithm $ 1 $ to identify cell-edge user signals. Exploiting the fact that communication signals are uncorrelated in time, and thus two copies of the same signal shifted by even one symbol are already uncorrelated, common user signals cannot be extracted via CCA if the $ \widetilde{T} $ symbols are misaligned. The correlation coefficient, $ \rho_n $, associated with each pair of canonical directions of $ \overline{\bf Q}_1 $ and $ \overline{\bf Q}_2 $ will not be at its maximum in this case, $ \forall n \in \{1,\cdots,K_e\} $. Based on this key observation, we develop a CCA based algorithm that can re-synchronize and then recover cell-edge user signals.

Define $  \widetilde{\bf Y}_1(\tau_1) := \widetilde{\bf Y}_1(:,\tau_1:T+\tau_1 - 1) $ and $  \widetilde{\bf Y}_2(\tau_2) := \widetilde{\bf Y}_2(:,\tau_2:T+\tau_2 - 1) $. Furthermore, let us define a search window of size $ [w_L,w_R] $ symbols. Upon setting $ \tau_2 =  1 $,  the CSPU solves problem~\eqref{R13} using the signals $ \widetilde{\bf Y}_1(\tau_1) $ and $  \widetilde{\bf Y}_2(\tau_2) $ to obtain $\overline{\bf q}^{*}_1 := {\bf Q}_1^{*}(:,1) $ and $\overline{\bf q}^{*}_2 := \overline{\bf Q}_2^{*}(:,1)$. Then, the CSPU computes and stores the corresponding correlation coefficient $ \rho_1 $ between $ \widetilde{\bf Y}^T_1(\tau_1)\overline{\bf q}^{*}_1 $ and $ \widetilde{\bf Y}^T_2(\tau_2)\overline{\bf q}^{*}_2 $. If  $ \tau_2 \leq w_s $,  increment $ \tau_2 $ and repeat, where $ w_s:= w_R-w_L +1 $ is the window size. Finally, pick the value $ \tau^{*}_2 $ that gives the highest $ \rho_1 $. This procedure is summarized in Algorithm 2. Note that as the locations of the $ T $ symbols are not generally known, the value of $ \tau_1 $ is chosen such that $ \widetilde{\bf Y}_1(\tau_1) $ includes a sufficient part of $ \overline{\bf Y}_1 $. This is guaranteed as long as $ w_s  << T$. 
\begin{algorithm} \label{A6}
	\SetAlgoLined
	\textbf{Input}:  $ \widetilde{\bf Y}_1 \in \mathbb{R}^{M_1 \times \widetilde{T}}  $, $ \widetilde{{\bf Y}}_2 \in \mathbb{R}^{M_2 \times \widetilde{T}} $ \\
	\textbf{Initialization}: $ \tau_2 :=  1 $ \\
	\While{$ \tau_2  \leq w_s$}{
			Compute $ {\bf \rho}_{1} $ after solving~\eqref{R13} using $ \widetilde{\bf Y}_1(\tau_1) $ and $ \widetilde{\bf Y}_2(\tau_2) $ \\
Store $(\tau_2,\rho_1)$ in a stack\\
			Set $ \tau_2 := \tau_2 + 1 $
		}
\textbf{Selection}: pick the $\tau_2$ corresponding to the highest $\rho_1$. 
	\caption{CCA SYNC}
	\label{A2}
\end{algorithm}   
     
Once we know the correct shift, $ \tau_d = \tau_2 -\tau_1 $, we can obtain  $ \overline{\bf Y}_\ell   $, for $ \ell \in \{1,2\} $, and use our proposed approach to identify cell-edge user signals using Algorithm 1. The computational complexity of Algorithm~\ref{A2} is that of solving for the principal component (canonical pair) of ~\eqref{R13} a number of times (equal to the search window).  The first canonical pair can be cheaply computed via a power iteration.  

\section{Experimental Results}\label{Simu}
To evaluate the performance of our proposed method, we consider a scenario with two hexagonal cells; each with radius $ R = 500 $ meters. Cell edge users are dropped randomly around the common edge between the two cells, i.e., the locations of cell-edge users were chosen randomly between $ 0.95R$ and $1.05R $. On the other hand, cell-center users are randomly dropped within distance $ zR $ from their serving BS, and we vary the value of $ z $ to see the effect of inter-cell interference on the proposed method. The transmitted power $ \beta_{kj} $ is set to $ 25 $dBm, $ \forall k,j $, i.e., power control is not employed. Furthermore, the transmitted sequence length $ T $ is fixed to $ 800 $. Additive white Gaussian noise is assumed with variance $ \sigma^2 $ so that the SNR is $ P_e / \sigma^2 $, where $ P_e $ is the average received power of cell-edge users. This enables us to see what values of SNR should cell-edge users have to achieve a specific BER. Furthermore, all results are averaged over $ 1000 $ channel realizations assuming different user locations in each realization. The uplink channel response vectors $ \{{\bf h}^H_{\ell kj }\} $ are modeled as
\begin{equation}
{\bf h}_{\ell kj}^H = \sqrt{\frac{1}{M_\ell}} \sum\limits_{n=1}^{L} \sqrt{\alpha_{\ell kj}^{(n)}}{\bf a}_r(\theta^{(n)})^H
\end{equation}
where $ L $ is the number of paths between the $ \ell $-th BS and the $ k $-th user in cell $ j $, $\forall \{\ell,j\} \in \{1,2\}~\text{and}~ k\in\{1,\cdots,K_\ell\} $. We use the path-loss model of the urban macro (UMa) scenario from the 3GPP $ 38.901 $ standard to compute the complex path gain $ \alpha_{\ell kj}^{(n)} $, $ \forall n,\ell,j,k $. Cell-center users were allowed to have a line of sight (LOS) path according to the LOS probability in the 3GPP $ 38.901 $ standard, however, all cell-edge users were non-LOS. The term $ {\bf a}_r(.) $ is the array response vector at the BS, and $ \theta^{(n)} \sim \mathcal{U}[-\pi,\pi]$ denotes the azimuth angle of arrival of the $ n $-th path. Assuming the BS is equipped with a uniform linear array, then
\begin{equation}
{\bf a}_r(\theta) = [1,\exp^{ikd\cos(\theta)}, \cdots,\exp^{ikd(M-1)\cos(\theta)}]
\end{equation}
where $ k = 2\pi/\lambda$, $ \lambda $ is the carrier wavelength and $ d = \lambda/2$ is the spacing between antenna elements.     


In order to benchmark the performance of our proposed method, we adopted two baselines. First, we implemented zero-forcing successive interference cancellation (ZF-SIC) where the channels of the cell-center users were assumed to be perfectly known at their serving BSs. Specifically, each BS decodes its cell-center users signals using ZF, encodes them again and then subtracts them from its received signal. Afterwards, the residual signal from each BS will be passed to RACMA~\cite{van1997analytical} in order to identify cell-edge user signals. Finally, the bit error rate (BER) of the cell-edge users is computed at both BSs and the best of the two BERs is reported. Furthermore, in order to guarantee fairness, since we have assumed joint processing of the BSs received signals, both residual signals from both BSs are further sent simultaneously to RACMA and the resulting BER (from RACMA with ``double measurements'') is also reported. Second, we implemented maximum likelihood successive interference cancellation (ML-SIC) to decode and subtract cell-center users signals assuming perfect knowledge of their channels at their serving BS. However, since in the worst-case the ML detector requires enumeration over all possible sequences of cell-center users, we only used this baseline when the number of cell-center users is small. The CCA approach (first stage) was implemented in MATLAB, while the MATLAB codes written by A.-J. van der Veen~\cite{van1997analytical} were utilized for the RACMA (second stage) implementation. 
\begin{figure}
	\begin{minipage}{.5\textwidth}
		\centering
		\includegraphics[width=80mm,scale=2]{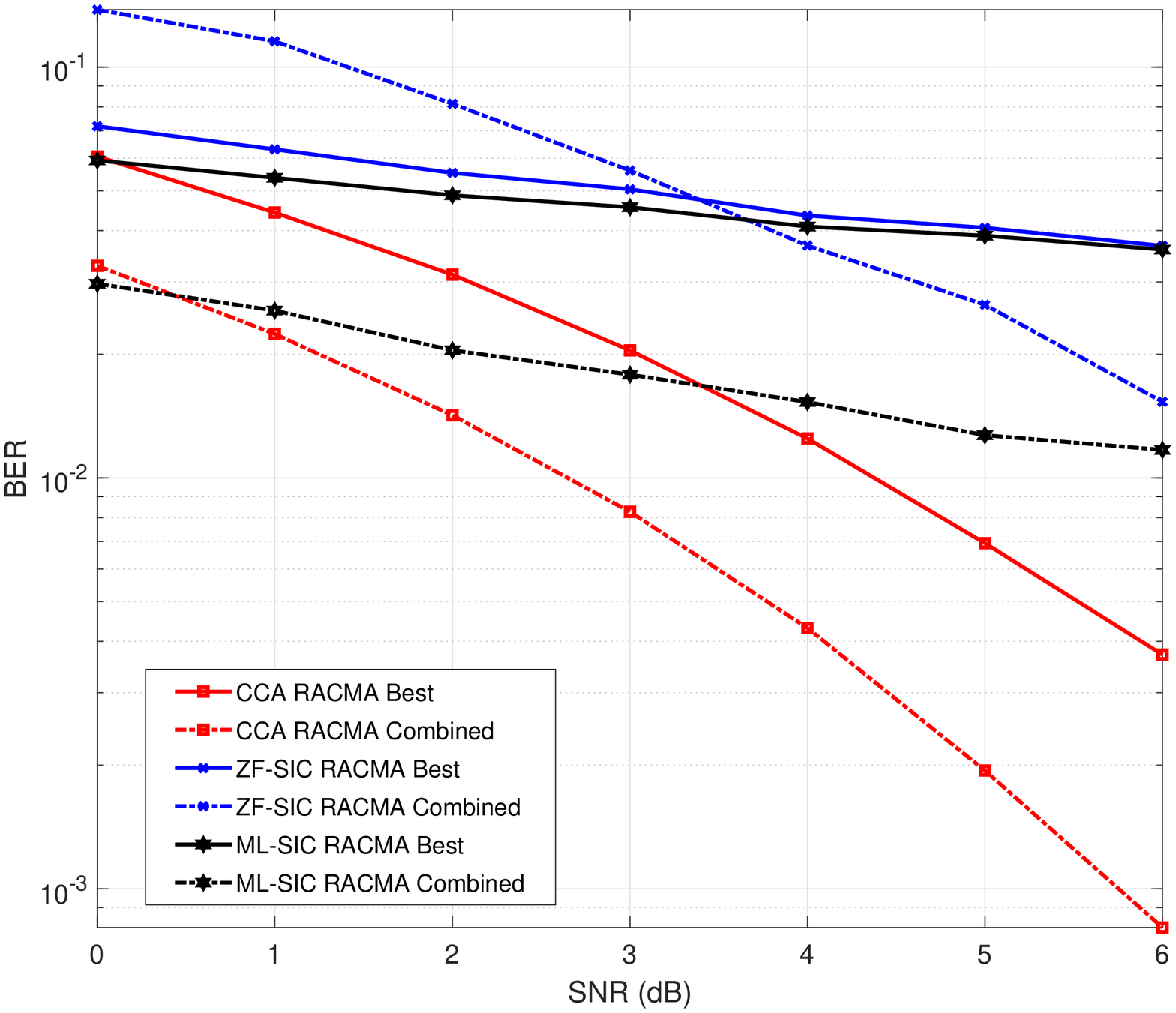}
		\caption{ BER vs. SNR of cell-edge users,  $ M_1 = M_2 =10 $, $ K_1 = K_2 = 8 $ and $ K_e=2 $, distance of cell-center users $ < $ 0.3R}
		\label{BER_SNR_d04_M10_K10_Kc2}
	\end{minipage}%
	\hspace{1cm}
	\begin{minipage}{.5\textwidth}
		\centering
		\includegraphics[width=80mm,scale=2]{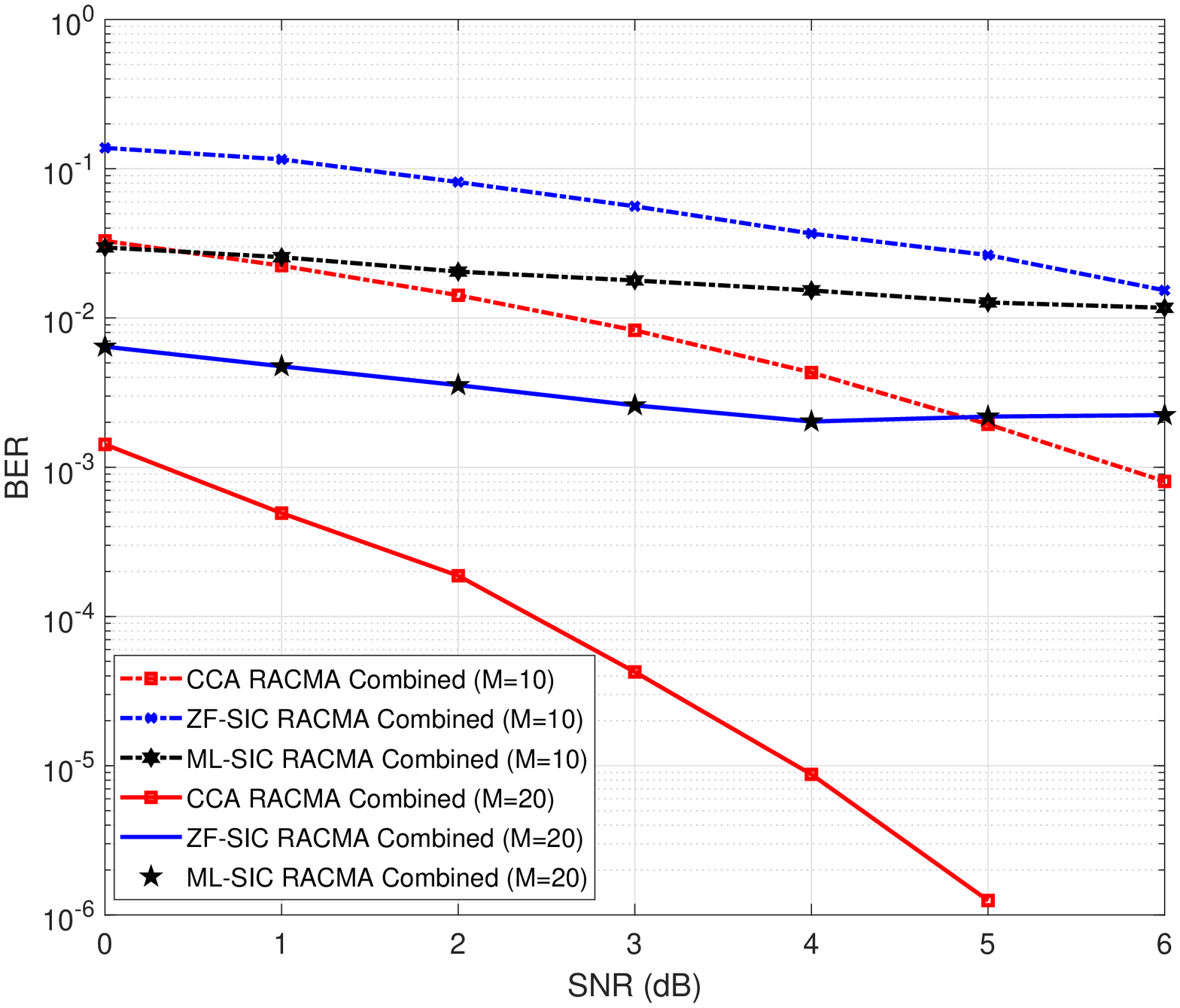}
		\caption{ BER vs. SNR of cell-edge users, $M_1 = M_2 = M$, $K_1 = K_2 = 8 $ and $ K_e=2 $, distance of cell-center users $ < $ 0.3R}
		\label{BER_SNR_d04_M1020_K10_Kc2}
	\end{minipage}%
\end{figure}

In a preliminary experiment, we consider a scenario with $ K_1 = K_2 = 8 $, $ M_1 = M_2 = 10 $, $ K_e = 2 $ and cell-center users are dropped randomly up to distance $ zR $, with $ z = 0.3 $. The numerical results for BER versus SNR of the cell-edge users is shown in Figure~\ref{BER_SNR_d04_M10_K10_Kc2}. It is obvious that our method significantly outperforms ZF-SIC and ML-SIC which assume perfect CSI of the cell-center users, whereas our method does not. For instance, more than one order of magnitude improvement using our CCA-RACMA method is observed at SNR$ = 6 $dB.

In order to see the effect of increasing the number of antennas on the performance of the proposed method, we considered the same setting of the previous experiment, however, we increased the number of antennas at each base station to $ 20 $, i.e., $ M_1 = M_2 = 20 $. Figure~\ref{BER_SNR_d04_M1020_K10_Kc2} shows that doubling the number of antennas at each base station improves the BER of cell-edge users obtained by all methods. However, a significant improvement gap in the BER obtained by our ``blind'' method is observed compared to that of ZF-SIC and ML-SIC. For instance, while ZF-SIC achieves an order of magnitude reduction in BER with $ M = 20 $, CCA-RACMA attains more than three orders of magnitudes improvement in BER at $\text{SNR}  =5 \text{dB}$. Furthermore, Figure~\ref{BER_SNR_d04_M1020_K10_Kc2} shows that our approach does yield measurable BER when the SNR of cell-edge users exceeds 5dB. The reason is that CCA can aggressively suppress the inter-cell interference when the number of antennas exceeds the total number of users, as explained in Appendix~\ref{AP}.

\begin{figure}
	\begin{minipage}{.5\textwidth}
		\centering
		\includegraphics[width=80mm,scale=2]{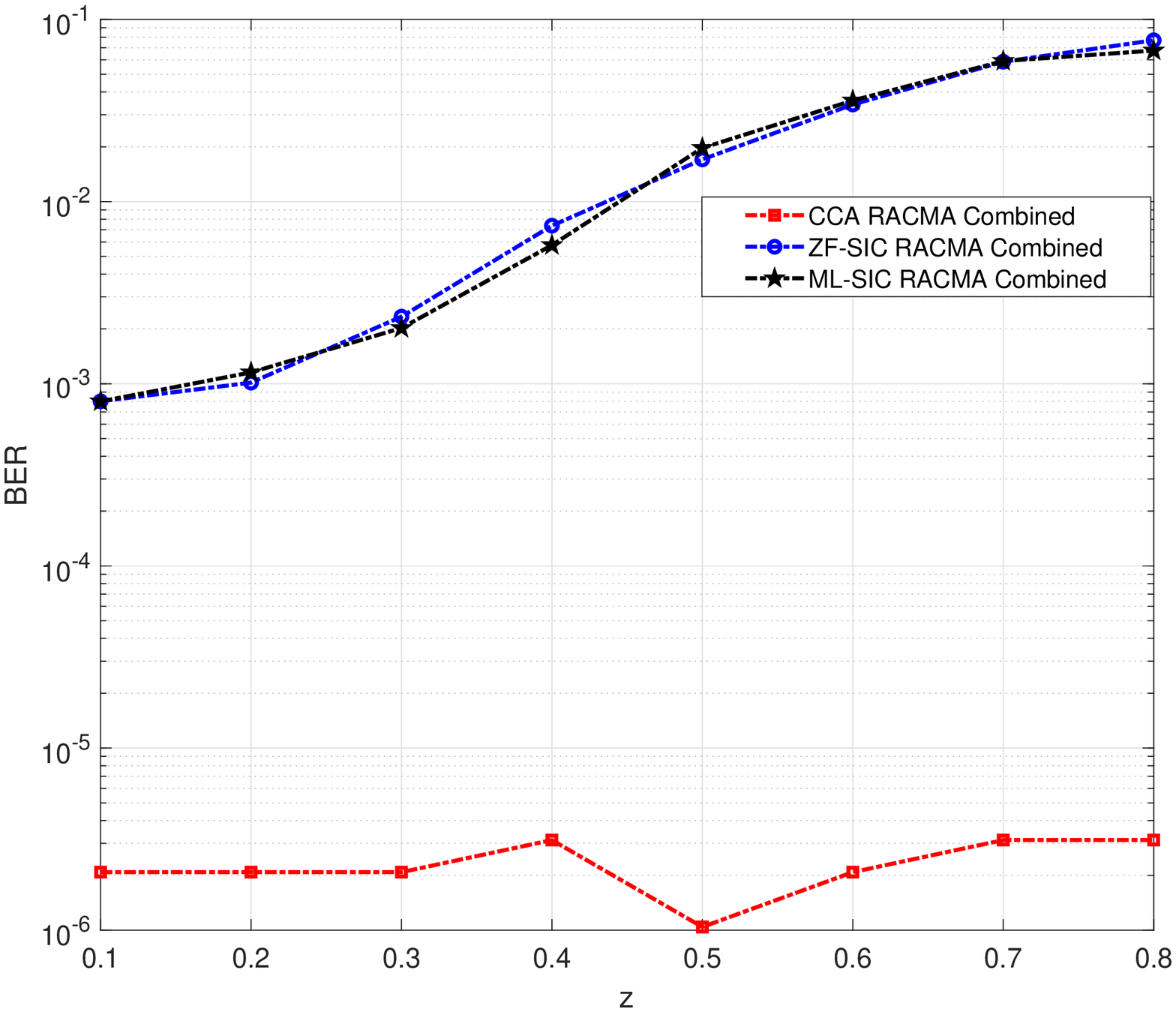}
		\caption{ BER vs. distance of cell-center users from their serving BS, with $ M_1 = M_2 =20 $, $ K_1 = K_2 = 8 $ and $ K_e=2 $, $ \text{SNR}=5\text{dB} $}
		\label{Inter-cell}
	\end{minipage}%
\hspace{1cm}
	\begin{minipage}{.5\textwidth}
		\centering
		\includegraphics[width=80mm,scale=2]{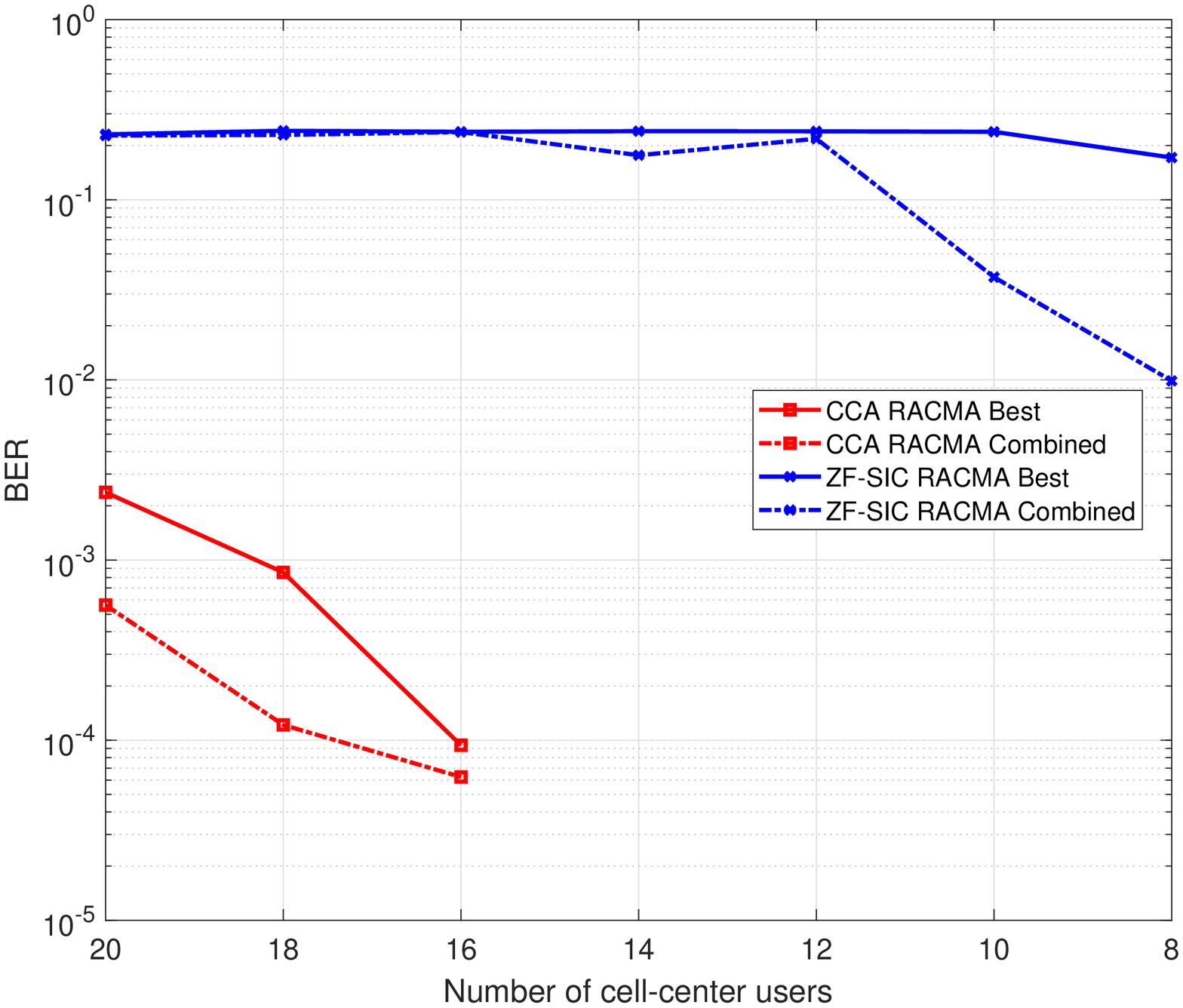}
		\caption{ BER vs. number of cell-center users at each BS, $M_1 = M_2 = 30 $, $ \text{SNR} = 4\text{dB} $ and $ K_e=2 $, distance of cell-center users   $ < $ 0.5R}
		\label{cell-center users}
	\end{minipage}%
\end{figure}


To test the effect of inter-cell interference, we vary the locations of cell-center users in their cell from $ 0.1R $ to $ 0.8R $, and for each setting we measure the BER attained by all methods at $ \text{SNR}= 5\text{dB} $. Figure~\ref{Inter-cell} demonstrates that the proposed CCA-RACMA approach still exhibits a favorable performance relative to that of ZF-SIC and ML-SIC. In particular, two orders of magnitude increase in the BER attained by ZF-SIC and ML-SIC is observed when the cell-center users are spread up to $ 0.8R $ compared to $ 0.1R $ from their serving BS, however, a very slight degradation in the performance of CCA-RACMA is observed, even for high spreads. Notice that, while the two baselines assume perfect knowledge of the cell-center user channels, this assumption becomes less realistic when ``cell-center'' users are in fact fully scattered throughout the cell. This therefore give a big advantage to the baselines over our method; notwithstanding, our method still works the best, even in this case. 

We now consider another experiment with $ M_1 = M_2 = 30 $, $ K_e = 2 $ and $ \text{SNR} = 5\text{dB} $. Assuming fixed user positions, we vary the number of cell-center users in each cell from $ 20 $ to $ 8 $, and for each given number of cell-center users we compute the BER of cell-edge users. In this experiment, all cell-center users are randomly dropped up to distance $ 0.5R $ from their serving BS. In Figure~\ref{cell-center users}, we observe that our proposed blind method can attain BER that is below the detectable threshold for this simulation when the number of cell-center users per cell is less than $ 16 $ while the ZF-SIC detector is severely affected by the cancellation errors from cell-center users. This shows how the proposed approach can handle dense scenarios, and hence, it is expected to work well in the case of multiple BSs (more than two).

\begin{figure}
	\begin{minipage}{.5\textwidth}
		\centering
		\includegraphics[width=80mm,scale=2]{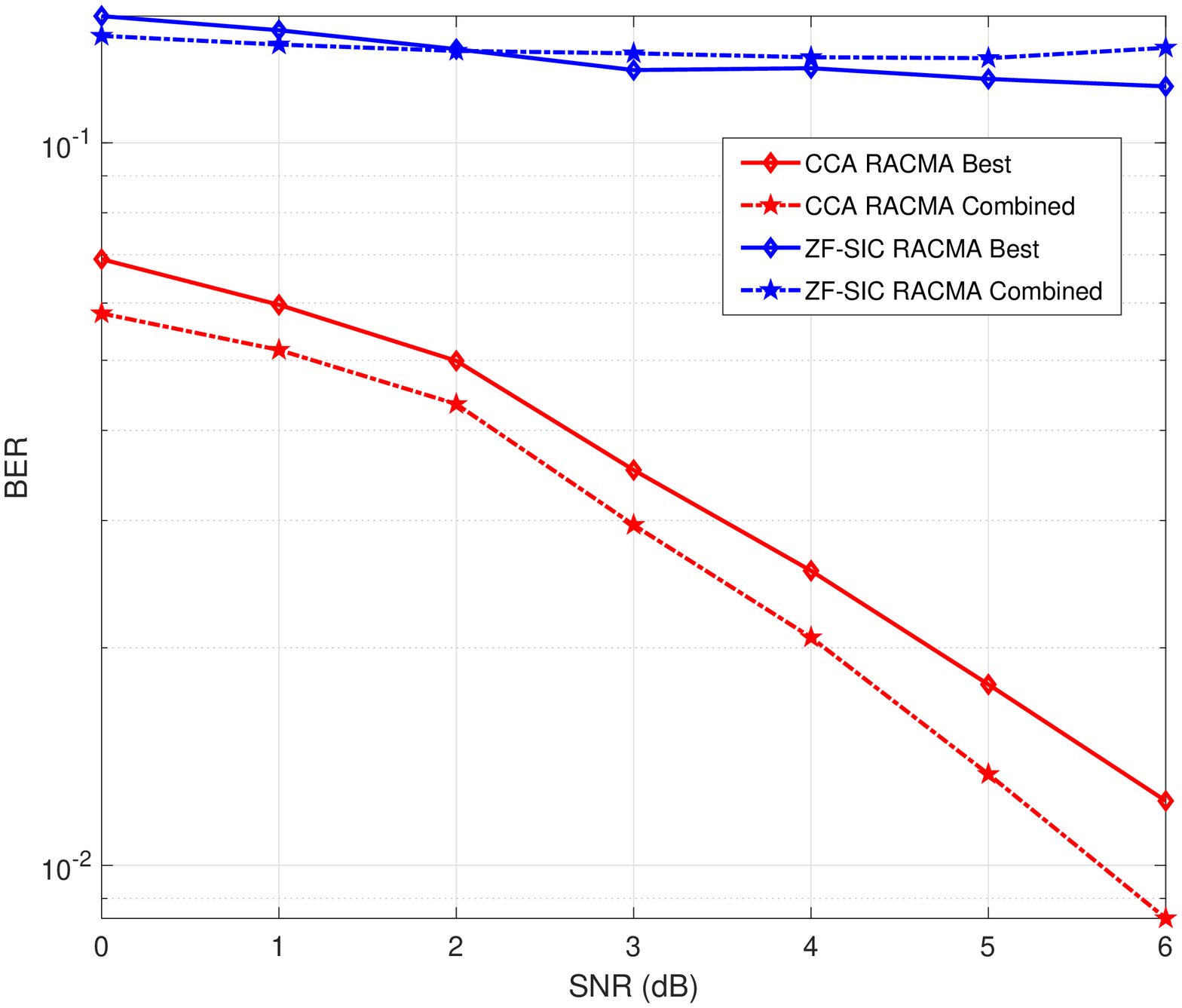}
		\caption{ BER vs. SNR, with $ M_1 = M_2 = 25 $, $ K_1 = K_2 = 15 $ and $ K_e=3 $, distance of cell-center users $ < $ 0.8R}
		\label{BER_SNR_d08_M25_K15_Kc3}
	\end{minipage}%
	\hspace{1cm}
	\begin{minipage}{.5\textwidth}
		\centering
		\includegraphics[width=80mm,scale=2]{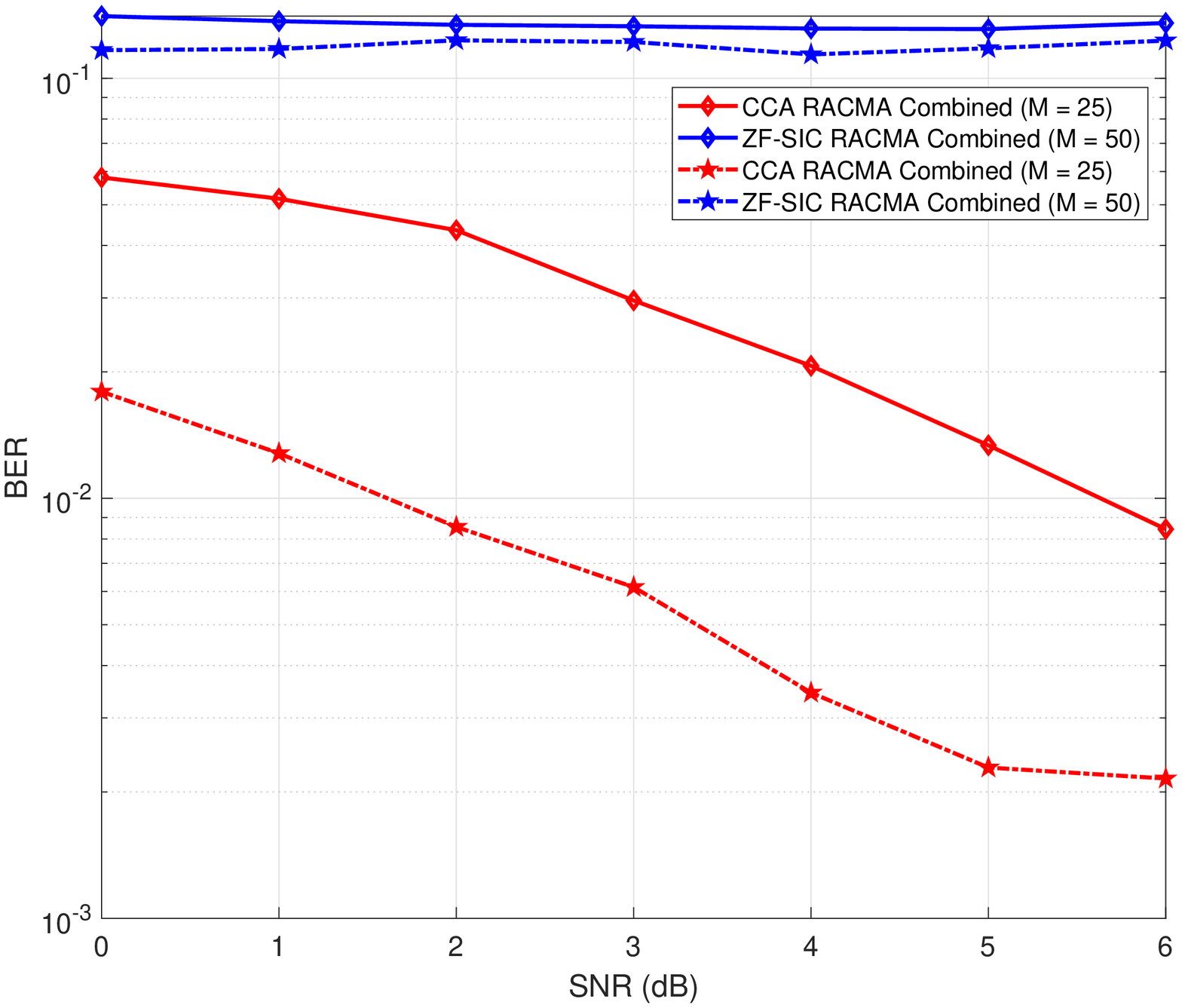}
		\caption{ BER vs. SNR of cell-edge users, with $ M_1=M_2=M $, $ K_1=K_2 = 15 $ and $ K_e=3 $, distance of cell-center users $ < $ 0.8R}
		\label{BER_SNR_d08_M2540_K15_Kc3}
	\end{minipage}%
\end{figure}

We next consider $ M_1 = M_2 = M = 25$, $K_1 = K_2 = 15$, $Ke = 3 $ and cell-center users are randomly located at distance less than $ 0.8R $ from their serving BS. As shown in Figure~\ref{BER_SNR_d08_M25_K15_Kc3}, jointly injecting more users (cell-center and cell-edge) and allowing them to be more spread, yields a noticeable degradation in the BER of cell-edge users achieved by all methods. This makes sense because, for ZF-SIC, there exists a higher chance that the detection performance of some cell-center users will be affected by the interference of cell-edge users resulting in cancellation errors from SIC. On the other hand, our method also exhibits some degradation in the performance because adding more users creates more intercell interference that can contaminate the common subspace estimated by CCA. However, our approach can still achieve much better performance to that obtained by ZF-SIC with perfect cell-center CSI. For example, our method still has more than an order of magnitude lower BER at different SNR values. 

We further simulate the previous scenario with double the number of antennas at each base station. As Figure~\ref{BER_SNR_d08_M2540_K15_Kc3} depicts, doubling the number of antennas at each base station yields an order of magnitude improvement in the BER of our method, while only slightly improving the BER of ZF-SIC.

\begin{figure}
	\centering
	\includegraphics[width=3in]{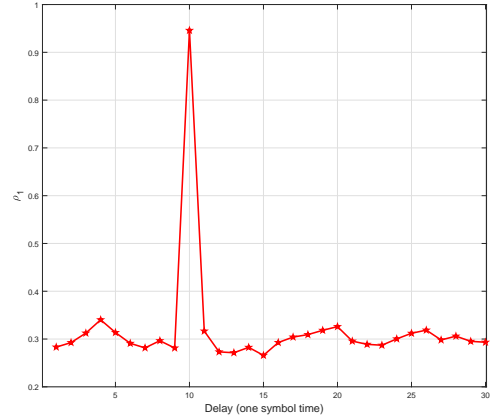}
	\caption{Correlation coefficient of the first pair-wise canonical component $ \rho_1 $ vs. delay}
	\label{sync_result}
\end{figure}

Finally, to show how CCA can still detect cell-edge user signals even when the received signals at the two BSs are not perfectly synchronized, we consider a scenario with $ K_1 = K_2 = 8 $, $ M_1 = M_2 = 10 $, $ K_e = 2 $, $ \text{SNR} = 3\text{dB} $, $ T = 800 $ and cell-center users are dropped randomly up to distance $ 0.5R $, and we assume that the received signal at the $ \ell $-th BS is  $ \widetilde{{\bf Y}}_\ell  \in \mathbb{R}^{M_\ell \times \widetilde{T}}$ , where $ \widetilde{T}$ was set to $ 830 $.  Then, we applied Algorithm~\ref{A2} on $ \widetilde{\bf Y}_1 $ and  $\widetilde{\bf Y}_2$, and observed the correlation coefficient of the first pair-wise canonical components as a function of the relative shift.  Figure~\ref{sync_result} shows how CCA can clearly identify the correct delay, and hence, detect cell-edge user signals as explained before. Clearly when the BS signals are not synchronized, there is no meaningful common subspace -- even the first pair of canonical components exhibits low correlation.  When we hit the correct delay, on the other hand, there are common components and the correlation coefficient $ \rho_1 $ is very high, as shown in Figure~\ref{sync_result}. 
  
\section{Conclusion and Future Work} \label{Conc} 
This paper has considered cell-edge user signal detection in the uplink of a multi-cell multi-user MIMO system. The goal is to design a detector that can reliably demodulate cell-edge user signals in the presence of strong intra-cell interference from users close to the base station, without resorting to power control or scheduling algorithms that throttle the cell-center user rates. This paper proposed a two-stage approach that leverages base stations cooperation to reliably identify cell-edge user signals at low SNR, without even knowing their channels. First, two-view CCA was brought in to estimate the subspace containing the cell-edge user signals shared by both base stations under the assumption that BS signals are synchronized. Then, an efficient analytical method called RACMA that guarantees the identifiability of binary signals from well-conditioned mixtures was utilized to extract the cell-edge user signals from the resulted mixture. We presented theoretical analysis of common subspace identifiability, in both ideal and realistic scenarios that include noise and inter-cell interference.   

Furthermore, we developed an algorithm that can identify cell-edge user signals in the case when  BS signals are not fully synchronized at the CSPU. Extensive simulations using a realistic path-loss model were carried out to show the superiority of the proposed learning-based method. It was shown that our blind CCA method achieves more than an order of magnitude improvement in the cell-edge user BER compared to the `oracle' zero forcing  and maximum likelihood cell-center multiuser detection followed by interference cancellation of the cell-center users before detecting the cell-edge users.  

In the future, it is interesting to study the more general setting of the problem that considers (possibly many) more than two `entangled' base stations. This introduces much more interference on cell-edge users that renders the problem much more challenging. In such scenarios, one can resort to generalized canonical correlation analysis to detect unknown cell-edge users whose signals are received with relatively equal power at multiple BSs. More importantly, one can investigate how many base stations should cooperate to provide the best detection performance for cell-edge users, and construct a performance-complexity blueprint. In addition, it is crucial to develop an efficient algorithm that can resolve the synchronization issue in the case of cooperation amongst a large number of BSs.


	\appendices
\section{Proof of Proposition 1}\label{AP}
In order to see how CCA can identify cell-edge user signals in the noisy and inter-cell interference case, let us first rewrite the received signal at the $ \ell $-th BS as
\begin{equation}\label{P1}
{\bf Y}_\ell = \sqrt{\beta_p}{\bf H}_{\ell p_\ell}{\bf S}_{p_{\ell}}^T + \sqrt{\beta_e}{\bf H}_{\ell c}{\bf S}^T_c + \sqrt{\beta_f}{\bf H}_{\ell p_\ell}{\bf S}_{p_{j}}^T+{\bf W}_\ell   
\end{equation}
where $ \ell,j \in \{1,2\} $ and $ j \neq \ell $. The terms $ \beta_p,~\beta_e $ and $ \beta_f $ represent the received power of the transmitted signal from the $ \ell $-th cell-center user to the $ \ell $-th BS, the received power of the transmitted signal from the cell-edge user to the $ \ell $-th BS, and the received power of the transmitted signal from the $ j $-th cell-center user to the $ \ell $-th BS, respectively,  for $ \ell \neq j $. Note that $ \beta_p $  integrates the transmitted power and the path-loss from each cell-center user to its serving BS, and the same applies for $ \beta_e $ and $ \beta_f $. For the sake of simplicity, we assume here that the cell-edge user signal is received with equal power at both BSs. Define $ {\bf H}_1 := [{\bf H}_{1c},{\bf H}_{1 p_1},{\bf H}_{1 p_2}] \in \mathbb{C}^{M_1 \times K_s}$ and $ {\bf H}_2 := [{\bf H}_{2c},{\bf H}_{2 p_1},{\bf H}_{2 p_2}] \in \mathbb{C}^{M_2 \times K_s}$, where $ K_s = K_1 + K_2 $. Each entry of $ {\bf H}_\ell $ represents the small scale fading between each user and the antennas in BS $ \ell $, and it is assumed to be independent complex zero-mean i.i.d. Gaussian random variable with variance $ 1/M_\ell $. It is further assumed that the number of antennas at the $ \ell $-th BS is greater than $ K_s $.

Let us first consider a simple scenario with two cell-center users (one at each BS) and one cell-edge user located at the common edge between the two BSs. We will now compute the cross- and auto-correlation matrices  $ {\bf R}_{{\bf y}_1{\bf y}_2} $, $ {\bf R}_{{\bf y}_\ell{\bf y}_\ell} $ as follows. First, we write~\eqref{P1} in more compact form by defining $ {\bf P}_1 = \text{Diag}([\beta_e,\beta_p,\beta_f]) $, $ {\bf P}_2 = \text{Diag}([\beta_e,\beta_f,\beta_p]) $ and $ {\bf B}  =  [{\bf s}_e,{\bf s}_{p_1},{\bf s}_{p_2}] $, where $ {\bf D}= \text{Diag}({\bf d}) $ is a diagonal matrix with the vector $ {\bf d} $ on its diagonal. The received signal at the $ \ell $-th BS can be written as
\begin{equation}\label{Ap5}
{\bf Y}_\ell = {\bf H}_\ell{\bf P}_\ell^{1/2}{\bf B}^T + {\bf  W}_\ell
\end{equation}
Since  the cross correlation matrix, $ {\bf R}_{{\bf y}_1{\bf y}_2} $, is given by  $ \frac{1}{T}{\bf Y}_1{\bf Y}_2^H $, then it follows that $ {\bf R}_{{\bf y}_1{\bf y}_2} $ is given by
\begin{equation}\label{Ap6}
    \begin{aligned}
        {\bf R}_{{\bf y}_1{\bf y}_2} &= \frac{1}{T} ({\bf H}_1{\bf P}_1^{1/2}{\bf B}^T+{\bf W}_1)({\bf H}_2{\bf P}_2^{1/2}{\bf B}+{\bf W}_2)^H \\
        &={\bf H}_1{\bf P}_{12}{\bf H}^H_2
    \end{aligned}
\end{equation}
where $ {\bf P}_{12} = ({\bf P}_2{\bf P}_1)^{1/2} $. Note that, in~\eqref{Ap6}, in addition to the assumption that  $ \frac{1}{T}{\bf B}^T{\bf B} = {\bf I} $, we exploited the fact that, for large $ T $, $\frac{1}{T} {\bf W}_\ell{\bf W}^H_j \approx 0$ and $\frac{1}{T} {\bf B}^T{\bf W}^H_j \approx 0$, for $ j,\ell \in \{1,2\} $. Similarly, the auto-correlation matrix of the received signal of the $ \ell $-th BS can be expressed as     
\begin{equation}\label{Ap7}
{\bf R}_{{\bf y}_\ell{\bf y}_\ell} =  {\bf H}_\ell{\bf P}_\ell{\bf H}_\ell^H + \sigma^2{\bf I}
\end{equation}
Now, we substitute with~\eqref{Ap6} and \eqref{Ap7} in~\eqref{R9} to obtain
\begin{equation}\label{Ap8}
\begin{aligned}
      {\bf H}_1{\bf P}_{12}{\bf H}_2^H ({\bf H}_2{\bf P}_2{\bf H}_2^H + \sigma^2{\bf I})^{-1} &{\bf H}_2{\bf P}_{12}{\bf H}_1^H{\bf q}_1  \\
     & = \lambda^2( {\bf H}_1{\bf P}_1{\bf H}_1^H + \sigma^2{\bf I}){\bf q}_1
\end{aligned}
\end{equation}
which can be equivalently written as 
\begin{equation}\label{Ap9}
 \begin{aligned}
    {\bf H}_1{\bf \Gamma}_{12}{\bf H}_2^H ({\bf H}_2{\bf \Gamma}_2{\bf H}_2^H + {\bf I})^{-1} &{\bf H}_2{\bf \Gamma}_{12}{\bf H}_1^H{\bf q}_1  \\
      & = \lambda^2( {\bf H}_1{\bf \Gamma}_1{\bf H}_1^H + {\bf I}){\bf q}_1
 \end{aligned}
\end{equation}
where $ {\bf \Gamma}_1 = \text{Diag}([\gamma_e,\gamma_p,\gamma_f]) $, $ {\bf \Gamma}_2 = \text{Diag}([\gamma_e,\gamma_f,\gamma_p]) $  and $  {\bf  \Gamma}_{12} = ({\bf \Gamma}_2{\bf \Gamma}_1)^{1/2} $, with $ \gamma_e  = \beta_e/ \sigma^2 $ be the received SNR of the cell-edge user, $ \gamma_p  = \beta_p/ \sigma^2 $ be the received SNR of each cell-center user at its serving BS,  and $ \gamma_f  = \beta_f/ \sigma^2 $ be the received SNR of each cell-center at the other (non-serving) BS. By left multiplying the two sides of~\eqref{Ap9} by $ {\bf H}^{\dagger}_1 $, we obtain
\begin{equation}\label{Ap10}
 \begin{aligned}
{\bf \Gamma}_{12}{\bf H}_2^H ({\bf H}_2{\bf \Gamma}_2{\bf H}_2^H + {\bf I})^{-1} &{\bf H}_2{\bf \Gamma}_{12}{\bf H}_1^H{\bf q}_1  \\
& = \lambda^2( {\bf \Gamma}_1{\bf H}_1^H + {\bf H}_1^\dagger){\bf q}_1
\end{aligned}
\end{equation}
By substituting with $ {\bf H}_1^\dagger = ({\bf H}_1^H{\bf H}_1)^{-1}{\bf H}_1^H $, and by letting $ {\bf v} = {\bf H}_1^H {\bf q}_1 $,~\eqref{Ap10} can be expressed as
\begin{equation}\label{Ap11}
{\bf \Gamma}_{12}{\bf H}_2^H ({\bf H}_2{\bf \Gamma}_2{\bf H}_2^H + {\bf I})^{-1} {\bf H}_2{\bf \Gamma}_{12}{\bf v} = \lambda^2( {\bf \Gamma}_1 +  ({\bf H}_1^H{\bf H}_1)^{-1} ){\bf v}
\end{equation}
By defining the matrix $ {\bf Z} := {\bf H}_2^H ({\bf H}_2{\bf \Gamma}_2{\bf H}_2^H + {\bf I})^{-1}{\bf H}_2 $, it then follows that $ {\bf Z} $ can be simplified as
\begin{subequations}\label{Ap12}
\begin{align}
      {\bf Z} &= {\bf H}_2^H ({\bf H}_2{\bf \Gamma}_2{\bf H}_2^H + {\bf I})^{-1}{\bf H}_2 \\
              &={\bf H}_2^{H} ({\bf H}_2^{\dagger}({\bf H}_2{\bf \Gamma}_2{\bf H}_2^H + {\bf I}))^{\dagger} \\
              &= {\bf H}_2^{H}({\bf H}_2^{H})^\dagger({\bf \Gamma}_2 + ({\bf H}_2^H{\bf H}_2)^{-1})^{-1} \\
              &= ({\bf \Gamma}_2 + ({\bf H}_2^H{\bf H}_2)^{-1})^{-1}
\end{align}
\end{subequations}
Note that in~(\ref{Ap12}b) and~(\ref{Ap12}c), we have exploited the following two properties of the pseudoinverse
\begin{enumerate}[label=P \arabic*.,itemindent=*]
	\item For any square matrix $ \bf A $, if $ {\bf A} $ is invertible, its pseudoinverse is its inverse, i.e., $ {\bf A}^{\dagger} = {\bf A}^{-1} $
	\item $ ({\bf BA})^{\dagger} = {\bf A}^{\dagger}{\bf B}^{\dagger} $
\end{enumerate}
By substituting with~(\ref{Ap12}d) in~\eqref{Ap11}, we obtain
\begin{equation}\label{Ap13}
         {\bf \Gamma}_{12}({\bf \Gamma}_2 + ({\bf H}_2^H{\bf H}_2)^{-1})^{-1}{\bf \Gamma}_{12}{\bf v}  = \lambda^2( {\bf \Gamma}_1 {+}  ({\bf H}_1^H{\bf H}_1)^{-1}){\bf v}
\end{equation}  
which can be equivalently expressed as
\begin{equation}\label{Ap14}
{\bf F}{\bf v} = \lambda^2 {\bf v}
\end{equation}
where $ {\bf F}{:=} ( {\bf \Gamma}_1 {+} ({\bf H}_1^H{\bf H}_1)^{-1})^{-1}{\bf \Gamma}_{12}({\bf \Gamma}_2 {+} ({\bf H}_2^H{\bf H}_2)^{-1})^{-1}{\bf \Gamma}_{12} $ is an $ K_s \times K_s $ matrix, and $ K_s = 3 $ for the particular scenario considered here. For ease of exposition, we will assume here that the number of antennas $ M_\ell $ is large enough so that $  ({\bf H}_\ell^H{\bf H}_\ell)^{-1} $ is approximately identity. Thus, matrix $ {\bf F} $ can be expressed as
	\[
{\bf F} := \left[
\begin{array}{ccc}\label{Ap15}
    (\frac{\gamma_e}{\gamma_e+1})^2&0&0\\
    0 & \frac{\gamma_f\gamma_c}{(\gamma_f + 1)(\gamma_c+1)} & 0 \\
    0 & 0 &\frac{\gamma_f\gamma_c}{(\gamma_f + 1)(\gamma_c+1)}
\end{array}
\right] \in \mathbb{R}^{K_s \times K_s},
\]
If each cell-center user is close to its serving BS, then  $ \gamma_f << 1 $ and $ \gamma_c >> 1 $. Therefore, the term $ \frac{\gamma_f\gamma_c}{(\gamma_f + 1)(\gamma_c+1)} $ will be approximately equal to $ \gamma_f $. Then, it can be easily seen that the maximum eigenvalue of the matrix $ {\bf F} $ is equal to $ (\frac{\gamma_e}{\gamma_e+1})^2 $ and the other two eigenvalues will be approximately equal to $\gamma_f$. Since the maximum eigenvalue of the matrix $ {\bf F} $ is nothing but the square of the correlation coefficient that is associated with the vectors $ {\bf Y}_1^T{\bf q}_1 $ and $ {\bf Y}_2^T{\bf q}_2 $. Then, it turns out that the maximum correlation coefficient is given by 
\begin{equation}\label{Ap16}
\rho_{\max} = \frac{\gamma_e}{\gamma_e +1 }
\end{equation} 
Now, we need to compute the eigenvectors $ {\bf q}_1 $ and $ {\bf q}_2 $. Since the maximum eigenvector of the diagonal matrix $ {\bf F} $ is given by
\begin{equation}\label{Ap17}
        {\bf v} = [\pm 1,0,0]^T
\end{equation}
the  eigenvector $ {\bf q}_1 $ can be obtained by solving the following system of linear equations
\begin{equation}\label{Ap18}
{\bf v} = {\bf H}_1^H {\bf q}_1
\end{equation}
Without loss of generality, we can let $ {\bf q}^{*}_1  =  {\bf H}_{1}({\bf H}_1^H{\bf H}_1)^{-1}{\bf v} $. The reason is that we can always find two components to the vector $ {\bf q}_1^{*} $; one in the subspace spanned by $ {\bf H}_1 $ and one orthogonal to it, however, the latter will vanish after multiplication with $ {\bf H}_1^{H} $. By substituting with ${\bf q}^{*}_1 $ in~\eqref{R8}, it can be easily proved that the corresponding canonical component of the second view  ${\bf q}_2^{*} = {\bf H}_2({\bf H}_2^H{\bf H}_2)^{-1}{\bf v} $. 
Define $ \hat{\bf s}_{\ell c} := {\bf Y}_\ell^H {\bf q}_\ell^{*}$ and substitute with $ {\bf q}_\ell^{*} $, we get the following
\begin{equation}\label{Ap20}
\hat{\bf s}_{\ell c} = \sqrt{\beta_e}c{\bf s}_{c} + {\bf n}_{\ell}
\end{equation}
where $ {\bf n }_\ell = {\bf W}_{\ell}^{H}{\bf q}_\ell^{*}  \in \mathbb{C}^{T}$ and $ c = \pm 1 $. This means that, in the case of single cell-edge user, the proposed detector can efficiently recover cell-edge user signals at low SNR even in the presence of inter-cell interference.

The generalization to $ K_e > 1$ and $ K_\ell- K_e > 1$ now follows directly. In that case, the matrix  $ F $ 
will have the vector $ {\bf f} \in \mathbb{R}^{K_s}$ on its diagonal, where 
\[ {\bf f}(j) =
\begin{cases}
(\frac{\gamma_{e_j}}{\gamma_{e_j}+1})^2,       & \quad  j \in \{1,\cdots,K_e\}\\
\frac{\gamma_{f_j}\gamma_{p_j}}{(\gamma_{f_j} + 1)(\gamma_{p_j}+1)}  & \quad j \in \{K_e + 1,\cdots,K_s\}
\end{cases}
\]
Assume that $ \gamma_{f_j} << 1, \forall j \in \{K_e +1,\cdots,K_s\}$. Then it can be easily seen that the largest $ K_e $ eigen vectors are the first $ K_e $ columns of an $ K_s \times K_s $ identity matrix. Upon letting $ {\bf V} = {\bf I}(:,1:K_e) $, the optimal solution $ {\bf Q}^{*}_\ell = {\bf H}_{\ell}({\bf H}_\ell^H{\bf H}_\ell)^{-1}{\bf V}{\bf M}_\ell$, where $ {\bf M}_\ell $ is any $ K_e \times K_e $ non singular matrix that satisfies the $ \ell $-th orthonormality constraint in~\eqref{R13}. Define $ \hat{\bf S}_{\ell c} := {\bf Y}^H_\ell {\bf Q}^{*}_\ell$ and substitute in~\eqref{P1}, we obtain
\begin{equation}\label{Ap21}
\hat{\bf S}_{\ell c} = {\bf S}_{c}{\bf P}^{1/2}_c{\bf M}_\ell + {\bf N}_{\ell}
\end{equation} 
where $ {\bf P}_c = \text{Diag}([\beta_{e_1},\cdots,\beta_{e_{K_e}}])$, and $ {\bf N}_{\ell}  = {\bf W}_{\ell}^{H}{\bf Q}_\ell^{*} $. Note that, after obtaining $ \hat{\bf S}_{\ell c}  $, we pass it to RACMA to identify the cell-edge user signals $ {\bf S}_c $.


\bibliographystyle{IEEEtran}
\bibliography{IEEEabrv,refrences}

\end{document}